\documentclass[11pt]{article}
\usepackage{moriond,epsfig}

\def\be{\begin{equation}}
\def\ee{\end{equation}}
\def\bea{\begin{eqnarray}}
\def\eea{\end{eqnarray}}

\begin{document}
\vspace*{0cm}
\title{SEARCHES FOR NEW PHYSICS USING DIJET EVENTS AT THE LHC}

\author{ M. CARDACI \\
on behalf of the ATLAS and CMS Collaborations }
\address{Universiteit Antwerpen, campus Groenenborger, \\
Physics Department / Building U, \\
Groenenborgerlaan 171, B-2020 Antwerpen, Belgium}

\maketitle\abstracts{
We present a concise review of the plans of the ATLAS and CMS Collaborations to search for new physics using dijet events.
The inclusive cross section as a function of jet p$_{T}$, the dijet mass distribution and the dijet ratio's CMS techniques are presented together with their potential of discovery with a focus on the integrated luminosities of 10, 100 and 1000 pb$^{-1}$. Analogously, the inclusive jet and angular distribution searches in the dijet channel from ATLAS are presented together with the potential of discovery given in terms of needed integrated luminosity. }

\section{Introduction}

Inclusive dijet production is the dominant LHC process. To lowest order it arises from the 2 $\rightarrow$ 2 scattering of partons. Inclusive jets and dijets both originate from this Standard Model (SM) scattering.
Dijet resonances from theories such as Technicolor \cite{Lane:2002sm}, Grand Unified Theory \cite{Eichten:1984eu}, Superstrings \cite{Hewett:1988xc}, Compositeness \cite{Baur:1987ga}, Extra Dimensions \cite{Randall:1999ee} and Extra Color \cite{Bagger:1987fz} \cite{Simmons:1996fz}, and Contact Interactions \cite{Eichten:1983hw} \cite{Lane:1996gr} \footnote{Both ATLAS and CMS use the CI lagrangian term: L$_{qqqq}$ = A (g$^{2}/$2$\Lambda^{2}_{LL}$) (\={q}$_{L}\gamma^{\mu}$q$_{L}$)(\={q}$_{L}\gamma_{\mu}$q$_{L}$) where $\Lambda$ is the contact interaction scale, g$^{2}$ is by convention 4$\pi\alpha_{s}$ and with A = +1 (destructive interference sign). } (CI) are signals of New Physics (NP) which can be discovered with dijet events, thanks to the deviations they produce in the basic SM distributions.
Dijet resonances produce compelling signals, but require the incoming parton-parton collision energy to be close to the mass of the produced resonance, which must be kinematically accessible. CI's, i.e. from Quark Compositeness, produce more ambiguous signals, but come from an energy scale of NP, $\Lambda$, which can be significantly larger than the available collision energy.
The focus of this article will be on the CMS plans on analyzing the first available dijet data and the ATLAS results from simulation. Finally, a summary on the potential of discovery and exclusion of the two experiments will be given.

\section{CMS's dijet searches for NP}

\begin{figure}[hbtp]
  \begin{center}
    \resizebox{5.4cm}{!}{\includegraphics{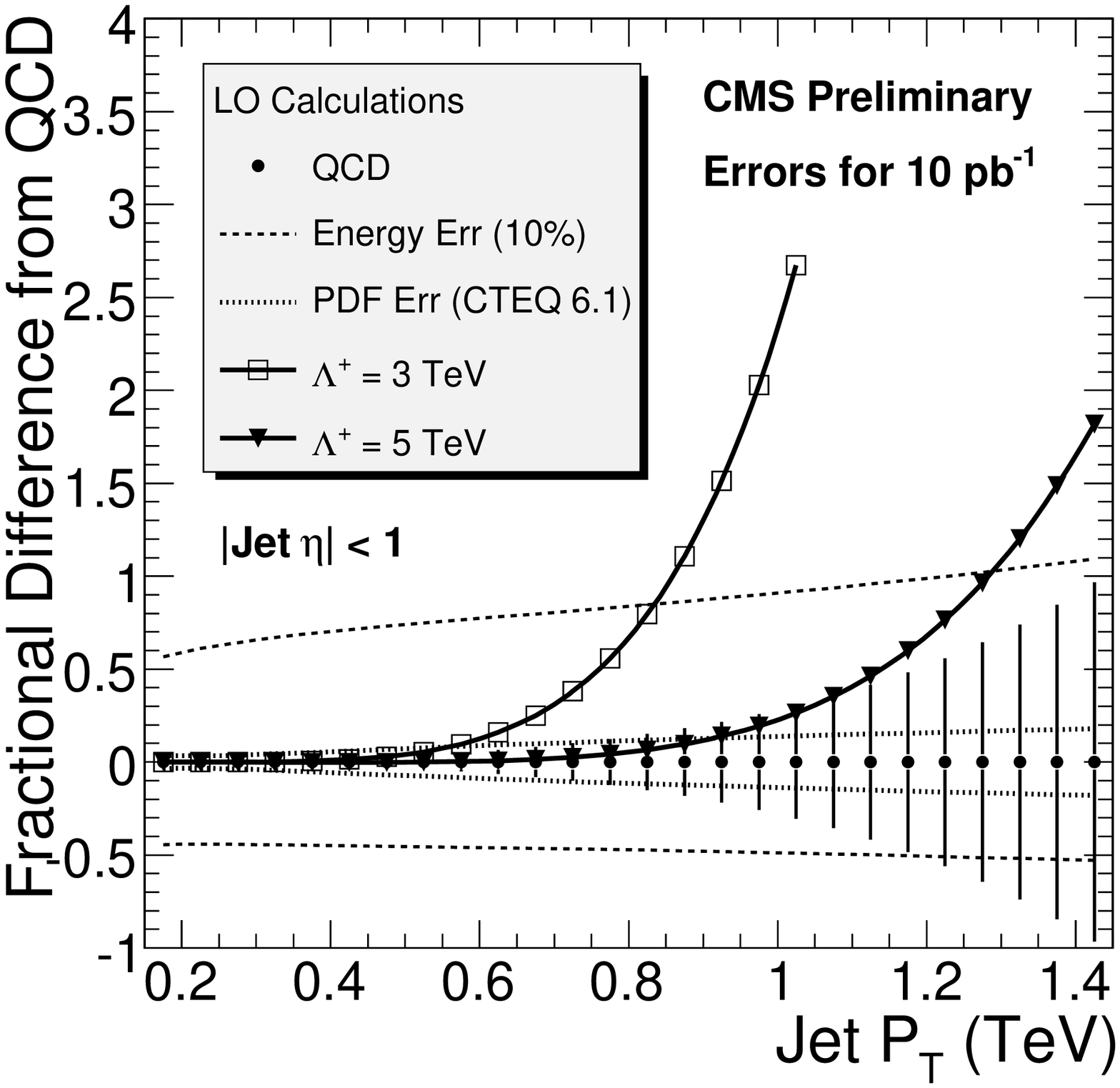}}
    \resizebox{6.4cm}{!}{\includegraphics{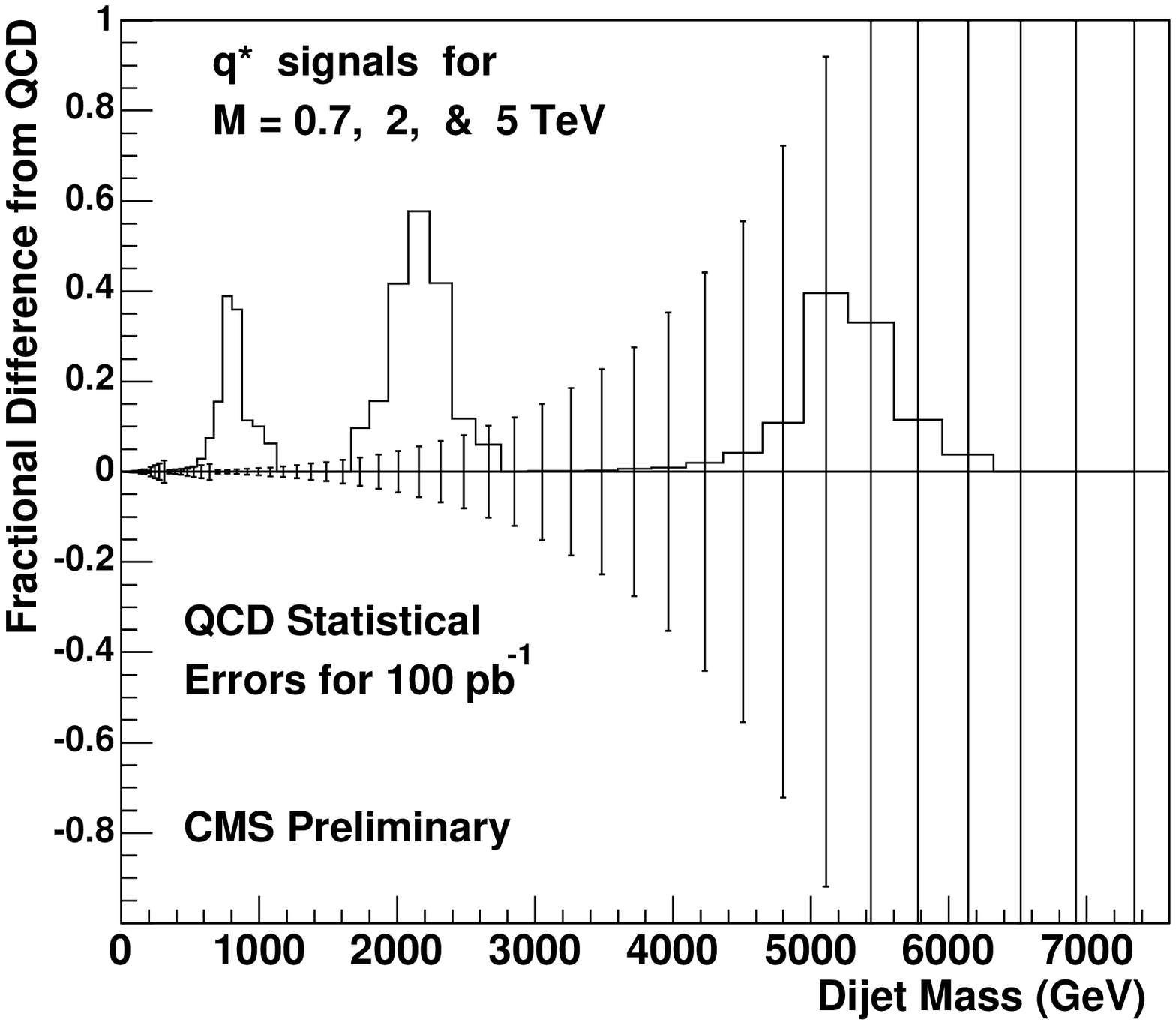}}
    \caption{Lefthand plot: Fractional deviation from the QCD background in the inclusive jet p$_{T}$ spectra of a CI for scale $\Lambda =$ 5 TeV (triangles) and scale $\Lambda =$ 3 TeV (squares) compared to the statistical error (vertical bars) and the energy scale and PDF uncertainties (dashed and dotted lines). Righthand plot: Fractional difference from the QCD background of a Z'-like spin 1 resonance, with the rate of q$^{*}$ resonance, for masses of 0.7, 2.0 and 5.0 TeV (solid line) and 100 pb$^{-1}$ of integrated luminosity, compared to the QCD statistical error (vertical bars).}
    \label{fig:Sensi}
  \end{center}
\end{figure}

CMS work presented here focus on sensitivities to NP for 10, 100 and 1000 pb$^{-1}$ of integrated luminosity.
CMS has developped 3 different techniques for its dijet searches, following the work started at the Tevatron \cite{Abe:1997hm} \cite{Abazov:2003tj}: the inclusive jet rate versus jet p$_{T}$ in the search of CI's, the dijet rate versus dijet mass in the search of resonances and the dijet ratio for both kind of NP signals.
The lefthand plot in Figure \ref{fig:Sensi} shows the fractional difference from the QCD background in the inclusive jet p$_{T}$ spectra of a CI for two different $\Lambda$ scales. A CI with scale $\Lambda =$ 3 TeV produces a clear rate deviation from the QCD expectation for jet p$_{T} >$ 1 TeV in the first 10 pb$^{-1}$, even taking into account a 10$\%$ energy scale uncertainty. The Tevatron has excluded $\Lambda$ up to 2.7 TeV \cite{Abbott:1998wh}, therefore, CMS has a clear possibility of discovery in the first data.
In the dijet rate versus dijet mass search, CMS can use a smooth fit or a QCD prediction to model the QCD background in order to estimate the sensitivity to NP shown in the righthand plot of Figure \ref{fig:Sensi} as fractional difference of a Z'-like spin 1 resonance, with the rate of an excited quark \cite{Baur:1987ga} (q$^{*}$) resonance, from the QCD background for 100 pb$^{-1}$ and compared to the statistical error. Such plot shows that a 2 TeV resonace has a convincing significance for only 100 pb$^{-1}$.

\begin{figure}[hbtp]
  \begin{center}
    \resizebox{5.9cm}{!}{\includegraphics{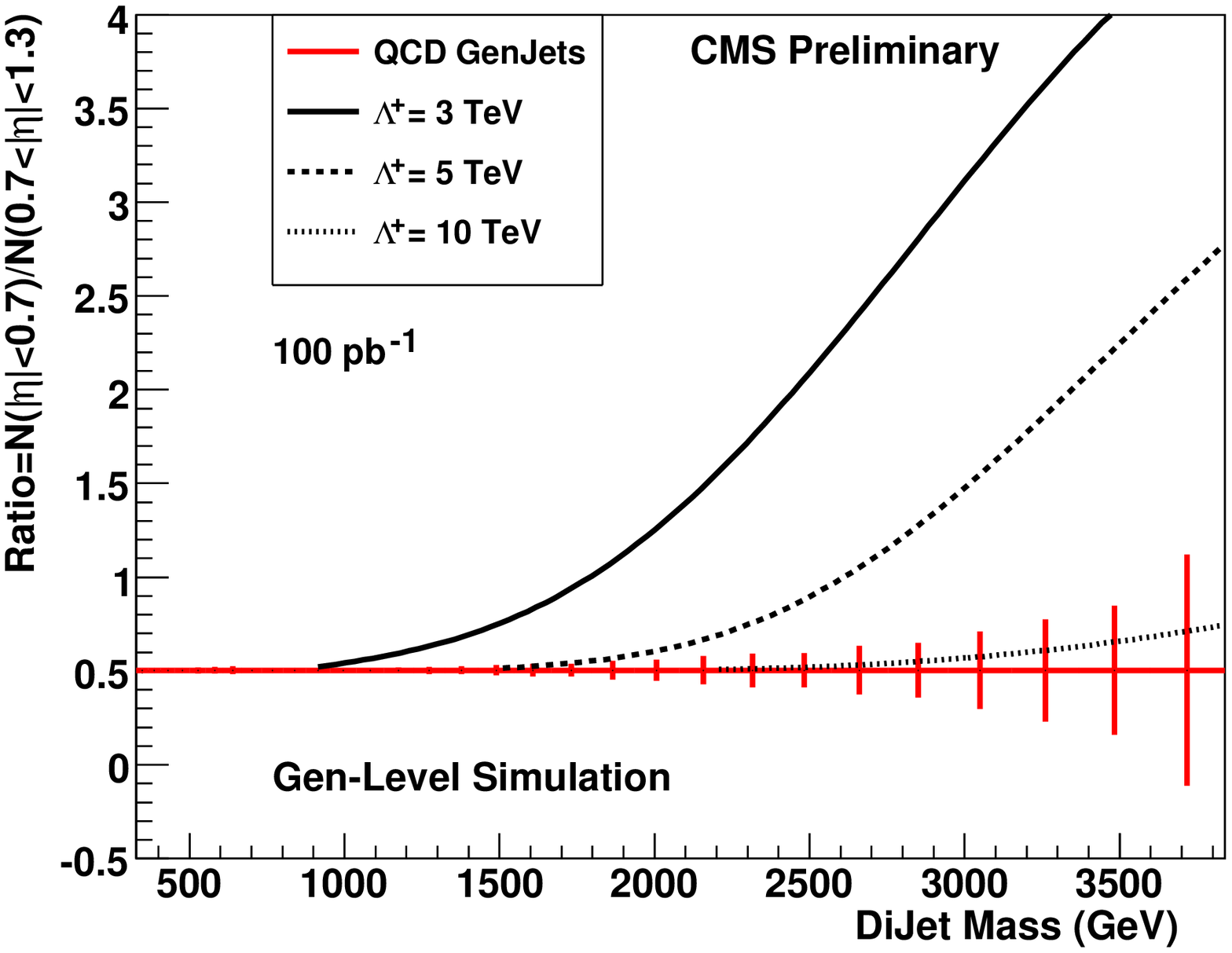}}
    \resizebox{4.8cm}{!}{\includegraphics{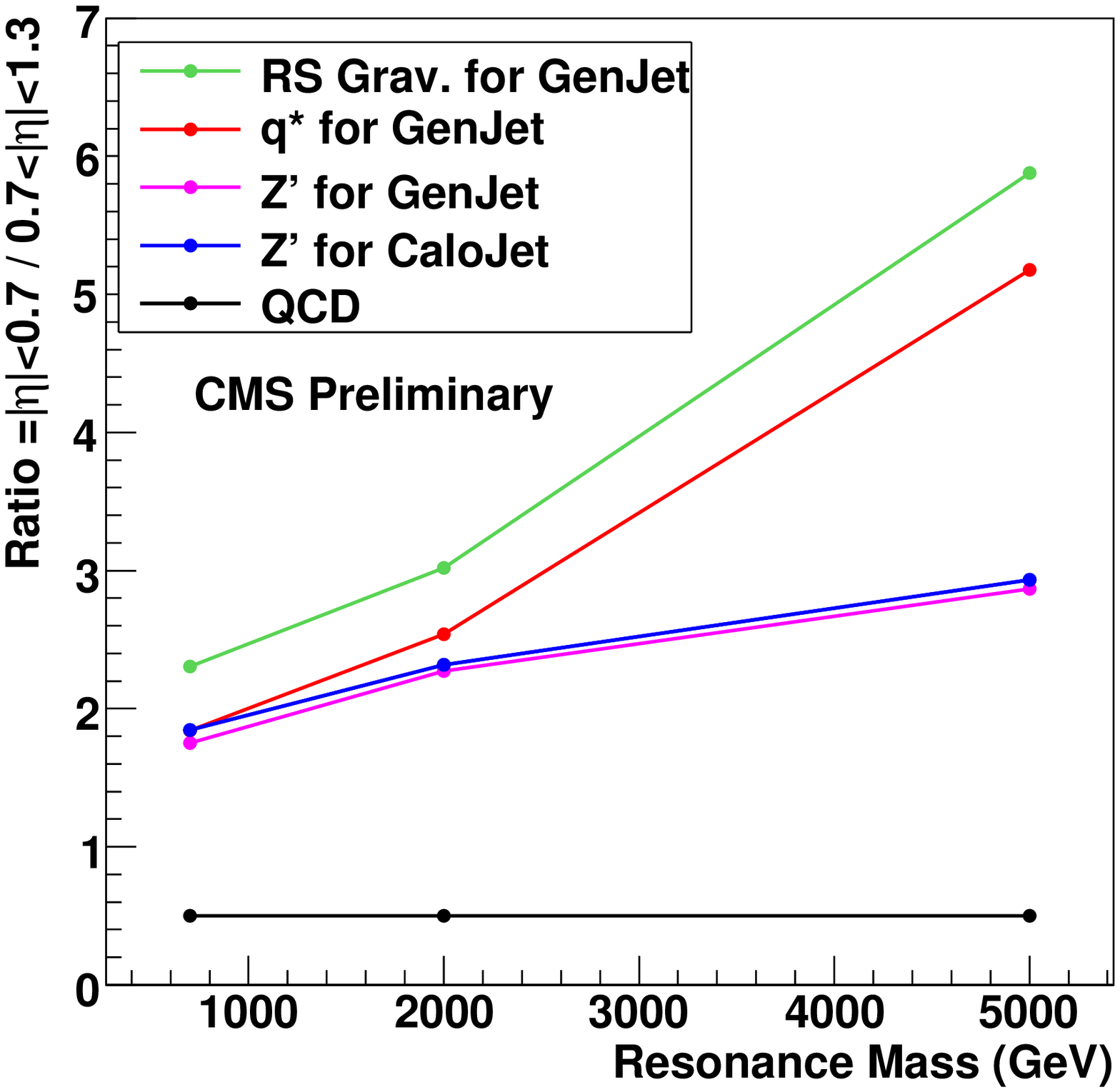}}
    \resizebox{4.8cm}{!}{\includegraphics{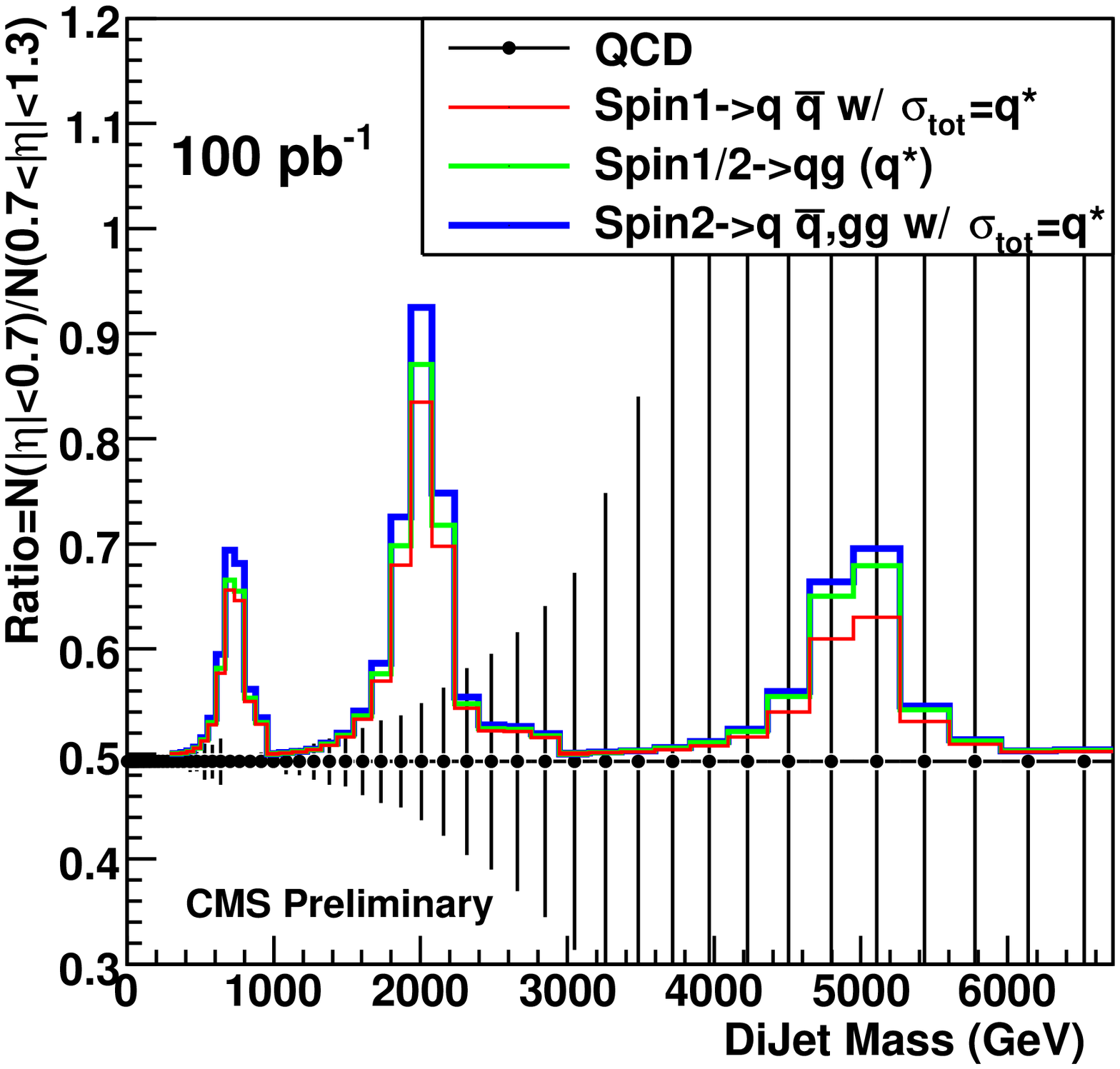}}
    \caption{Lefthand plot: Dijet ratio fractional difference from the QCD background of a CI for 3 different $\Lambda$ scales (solid, dashed and dotted lines) and 100 pb$^{-1}$ of integrated luminosity compared to the QCD statistical error (vertical bars). Central plot: Dijet ratio versus dijet mass for resonances of different spins. Righthand plot: Dijet ratio fractional difference versus dijet mass for resonances of different spins and cross section fixed to the q$^{*}$ cross section (solid line) compared to the QCD statistical error (vertical bars) for 100 pb$^{-1}$.}
    \label{fig:Sensi2}
  \end{center}
\end{figure}

\noindent Another technique is going to be used by CMS to discover NP: the dijet ratio. The dijet ratio is defined as the number of events in which each leading jet has $|\eta| <$ 0.7 divided by the number of events in which each leading jet has 0.7 $< |\eta| <$ 1.3. The two region cuts have been optimized on the sensitivity. The events in the numerator typically have values of cos $\theta^{*}$ (where $\theta^{*}$ is the angle between the intial and final state partons in the center of momentum frame) close to zero, so the numerator is sensitive to NP which tends to be relatively isotropic, pretty flat in cos $\theta^{*}$ . In contrast the events in the denominator typically have larger values of cos $\theta^{*}$, closer to 0.7. The denominator will mainly contain background from QCD dijets. The dijet ratio versus the dijet mass for the QCD dijets is constant at the value of 0.5 up to 6 TeV. CI affects the dijet ratio distribution at high mass. In the lefthand plot of Figure \ref{fig:Sensi2} is shown the fractional difference from the SM QCD of a CI, for 3 different $\Lambda$ scales and for 100 pb$^{-1}$. Detailed studies have demonstrated that CMS can discover a CI of scales $\Lambda =$ 4, 7 and 10 TeV with respectively 10, 100 and 1000 pb$^{-1}$. In the Figure \ref{fig:Sensi2} the central one is a pedagogical plot of the dijet ratio versus the dijet mass for resonances of different spins, showing the dependence of the dijet ratio with respect to the spin of the resonance. In Figure \ref{fig:Sensi2} the righthand plot shows the dijet ratio fractional deviation versus dijet mass for resonances of different spin and cross section fixed to the q$^{*}$ cross section. There is a convincing signal for a 2 TeV resonance with a q$^{*}$ cross section in 100 pb$^{-1}$ for all resonance spins considered. We expect that with sufficient luminosity the dijet ratio can be used to measure the resonance spin. Furthermore the dijet ratio has the advantage to reduce the systematic uncertainties due to the nature of the quantity itself measured: a ratio.
In addition, it is worth to notice that all the analysis techniques above mentioned will in the beginning have challenging detector and analysis issues to confront with. Detector noise, cosmic ray air showers and beam halo produce large missing transverse energy. A cut on the ratio MET$/ \Sigma E_{T}$ will be applied in order to filter out most of this unphysical background.

\section{ATLAS's dijet searches for NP}

To characterize the excess at high p$_{T}$ produced by a CI in the inclusive jet distribution, ATLAS uses the ratio between the number of jet having E$_{T}$ above a certain threshold and those having it below:

\begin{equation}
R(\Lambda) =\frac{N(E_{T} > E_{T}^{0})}{N(E_{T} < E_{T}^{0})}
\end{equation} \\

\noindent Similarly it does in dijet angular distribution defining a variable related to the pseudorapidity of the two leading jets:

\begin{equation}
\chi = e^{\eta_{1} - \eta_{2}} ;
R_{\chi}(\Lambda) = \frac{N(\chi < \chi^{0})}{N(\chi > \chi^{0})}
\end{equation} \\

\noindent In both cases the threshold is optimized (E$_{T}^{0}$ = 1100 GeV and $\chi^{0}$ = 2.8) and a variable to estimate the significance of the deviation from the Standard Model is given:

\begin{equation}
R_{dist} = \frac{R(\Lambda)-R(SM)}{\sqrt{\sigma_{R(\Lambda)}^{2} + \sigma_{R(SM)}^{2}}} ;
R_{1} = \frac{R_{\chi}(\Lambda)-R_{\chi}(SM)}{\sqrt{\sigma_{R_{\chi}(\Lambda)}^{2} + \sigma_{R_{\chi}(SM)}^{2}}}
\end{equation} \\

\noindent A detailed study on the inclusive jet p$_{T}$ distribution for 30 fb$^{-1}$ has shown that 1$\%$  uncertainty in the energy scale can hide a CI of scale  $\Lambda$ = 20 TeV. In the table \ref{tab} is shown that CI's of $\Lambda$ = 3, 5 and 10 TeV might be ruled out or verified with first tens of pb$^{-1}$. A first attempt to study PDF uncertainties with CTEQ6M based on NLO calculation and data fitted to DIS has shown a systematic error due to PDF uncertainties such as $\sigma_{R_{dist}}$(PDF) = 1.40. The error is comparable with the size of the significance for the 40 TeV scale CI, therefore the discovery of a CI of this $\Lambda$ scale is still unclear. \\

\begin{table}[h]
\caption{R values for $\mathcal{L}$ = 300 fb$^{-1}$ and $\mathcal{L}$ values to achieve R = 3 \label{tab}}
\vspace{0.4cm}
\begin{center}
\begin{tabular}{|c|c|c|c|c|c|}
\hline
$\Lambda$ (TeV) & 3 & 5 & 10 & 20 & 40  \\ \hline
R$_{dist}$ & 794 &  427  & 44 & 12 & 3.4  \\ \hline
R$_{1}$ & 2500 &  665  & 62 & 8.9 & 2.5  \\ \hline
$\mathcal{L}$(R$_{dist}$) & 4.3 pb$^{-1}$ & 15 pb$^{-1}$ & 1.4 fb$^{-1}$ & 19 fb$^{-1}$ & 234 pb$^{-1}$ \\ \hline
$\mathcal{L}$(R$_{1}$) & $<$ 1 pb$^{-1}$ & 6 pb$^{-1}$ & 0.7 fb$^{-1}$ & 34 fb$^{-1}$ & 426 pb$^{-1}$ \\ \hline
\end{tabular}
\end{center}
\end{table}

\noindent In the angular distribution searches ATLAS makes use of a jet p$_{T}$ cut of 350 GeV, optimized on the significance, and a $\Lambda$ dependent cut on the dijet mass. The most recent result on the angular distribution from ATLAS are shown in the table \ref{tab}. Also in this case discovery limits are presented and also in this case discovery of a CI of a $\Lambda$ scale of 40 TeV is still unclear. The systematic error due to PDF uncertainties is of the order of the significance: $\sigma_{R_{1}}$(PDF) = 0.88 and R$_{1}$($\Lambda$ = 40 TeV, 30 fb$^{-1}$) = 0.80. It is anyway preliminary to say anything concerning the possibility of discovery of a CI having $\Lambda$ = 40 TeV. \\

\section{Conclusions}

For CMS the inclusive jet p$_{T}$ analysis gives a convincing signal for a CI of scale $\Lambda$ = 3 TeV in 10 pb$^{-1}$. The rate versus dijet mass analysis gives a convincing signal for a 2 TeV q$^{*}$ with 100 pb$^{-1}$. With the dijet ratio technique CMS can discover or confirm a dijet resonance and eventually measure its spin. For ATLAS CI's of scale $\Lambda$ = 3, 5 and 10 TeV might be discovered or ruled out with first tens of pb$^{-1}$ of good data. $\Lambda$ = 20 TeV should be visible with larger integrated luminosity. Discovery of $\Lambda$ = 40 TeV CI is still unclear. Where comparable ATLAS and CMS have the same potential of discovery.

\section*{Acknowledgments}

Ackowledgements to the Dijet Group in CMS: A. Bhatti, B. Bollen, M. Cardaci, F. Chlebana, S. Esen, R. M. Harris, M. K. Jha, K. Kousouris, D. Mason and M. Zielinski and to the ATLAS collaborators: S. Ferrag and L. P\v{r}ibyl.


\section*{References}
\begin{thebibliography}{99}

\bibitem{Lane:2002sm}
  K.~Lane and S.~Mrenna,
  Phys.\ Rev.\  D {\bf 67}, 115011 (2003)
  [arXiv:hep-ph/0210299].

\bibitem{Eichten:1984eu}
  E.~Eichten, I.~Hinchliffe, K.~D.~Lane and C.~Quigg,
  Rev.\ Mod.\ Phys.\  {\bf 56}, 579 (1984)
  [Addendum-ibid.\  {\bf 58}, 1065 (1986)].

\bibitem{Hewett:1988xc}
  J.~L.~Hewett and T.~G.~Rizzo,
  Phys.\ Rept.\  {\bf 183}, 193 (1989).

\bibitem{Baur:1987ga}
  U.~Baur, I.~Hinchliffe and D.~Zeppenfeld,
  Int.\ J.\ Mod.\ Phys.\  A {\bf 2}, 1285 (1987).

\bibitem{Randall:1999ee}
  L.~Randall and R.~Sundrum,
  Phys.\ Rev.\ Lett.\  {\bf 83}, 3370 (1999)
  [arXiv:hep-ph/9905221].

\bibitem{Bagger:1987fz}
  J.~Bagger, C.~Schmidt and S.~King,
  Phys.\ Rev.\  D {\bf 37}, 1188 (1988).

\bibitem{Simmons:1996fz}
  E.~H.~Simmons,
  Phys.\ Rev.\  D {\bf 55}, 1678 (1997)
  [arXiv:hep-ph/9608269].

\bibitem{Eichten:1983hw}
  E.~Eichten, K.~D.~Lane and M.~E.~Peskin,
  Phys.\ Rev.\ Lett.\  {\bf 50}, 811 (1983).

\bibitem{Lane:1996gr}
  K.~D.~Lane,
  arXiv:hep-ph/9605257.

\bibitem{Abe:1997hm}
  See for instance: F.~Abe {\it et al.}  [CDF Collaboration],
  Phys.\ Rev.\  D {\bf 55}, 5263 (1997)
  [arXiv:hep-ex/9702004].

\bibitem{Abazov:2003tj}
  See for instance: V.~M.~Abazov {\it et al.}  [D0 Collaboration],
  Phys.\ Rev.\  D {\bf 69}, 111101 (2004)
  [arXiv:hep-ex/0308033].

\bibitem{Abbott:1998wh}
  B.~Abbott {\it et al.}  [D0 Collaboration],
  Phys.\ Rev.\ Lett.\  {\bf 82}, 2457 (1999)
  [arXiv:hep-ex/9807014].

\end{thebibliography}
\end{document}